\documentclass[12pt,preprint]{aastex}

\begin{document}

\title{Planet Formation with Migration}

\author{J. E. Chambers\altaffilmark{1}}
\altaffiltext{1}{Department of Terrestrial Magnetism, Carnegie Institution of Washington, 5241 Broad Branch Road NW, Washington DC 20015, USA; chambers@dtm.ciw.edu}

\begin{abstract}
In the core-accretion model, gas-giant planets form  solid cores which then accrete gaseous envelopes. Tidal interactions with disk gas cause a core to undergo inward type-I migration in $10^4$ to $10^5$ years. Cores must form faster than this to survive. Giant planets clear a gap in the disk and undergo inward type-II migration in $<10^6$ years if observed disk accretion rates apply to the disk as a whole. Type-II migration times exceed typical disk lifetimes if viscous accretion occurs mainly in the surface layers of disks. Low turbulent viscosities near the midplane may allow planetesimals to form by coagulation of dust grains. The radius $r$ of such planetesimals is unknown. If $r<0.5$ km, the core formation time is shorter than the type-I migration timescale and cores will survive. Migration is substantial in most cases, leading to a wide range of planetary orbits, consistent with the observed variety of extrasolar systems. When $r\sim 100$m and midplane $\alpha\sim 3\times 10^{-5}$, giant planets similar to those in the Solar System can form.
\end{abstract}


\section{Introduction}

In the core-accretion model, gas-giant planets begin life as solid cores that grow by sweeping up small planetesimals \citep{Inaba2003}. Cores $>10^{-5}M_\oplus$ grow oligarchically: each radial zone in a protoplanetary disk contains a single core and many planetesimals \citep{Kokubo1998,Thommes2003}. Perturbations from a core control the orbital distribution of nearby planetesimals which then determines the core's growth rate. Cores larger than Mars acquire extended atmospheres of nebular gas. Passing planetesimals are slowed by gas drag as a result, increasing the chance of capture \citep{InabaIkoma}. When a core reaches a critical mass $M_{\rm crit}\sim 10 M_\oplus$, a static atmosphere can no longer be supported, and the core accretes a massive gas envelope, becoming a gas-giant planet \citep{Hubickyj2005, Ikoma2000}. 

Cores must grow to $M_{\rm crit}$ before the disk gas disperses, which typically happens in a few Myr \citep{Haisch2001}. A core of mass $M$ sweeps up planetesimals at a rate given by
\begin{equation}
\left.\frac{dM}{dt}\right)_{\rm solid}=\left(\frac{2\pi\Sigma_{\rm solid}r_h^2}{P}\right)P_{\rm col}(e,i)
\label{eq-dmdt}
\end{equation}
where $P$ and $r_h$ are the core's orbital period and Hill radius, and $\Sigma_{\rm solid}$ is the local surface density of planetesimals \citep{Inaba2001}. The collision probability $P_{\rm col}$ depends on the relative velocity $v_{\rm rel}$ of passing planetesimals, which is a function of their mean eccentricity $e$ and inclination $i$ \citep{Inaba2001}. 

In the minimum-mass solar nebula (MMSN), the solid and gas surface densities are:
\begin{eqnarray}
\Sigma_{\rm solid}&=& \Sigma_0\,\eta_{\rm ice}\left(\frac{a}{\rm 5\ AU}\right)^x \nonumber \\
\Sigma_{\rm gas}&=&\Sigma_0\,\eta_{\rm gas}\left(\frac{a}{\rm 5\ AU}\right)^x
\label{eq-sigma}
\end{eqnarray}
where $\Sigma_0\,\eta_{\rm ice}\simeq 3$ g\,cm$^{-2}$ and $x\simeq-3/2$ \citep{Weidenschilling1977}.  Here, $a$ is the orbital radius; $\Sigma_0$ is the surface density of rocky material at 5 AU; $\eta_{\rm gas}\simeq 200$; $\eta_{\rm ice}=1$ inside the snow line, and $\eta_{\rm ice}>1$ outside the snow line. Most models for the formation of Jupiter use mass-enhanced nebulae, with $\Sigma_0\,\eta_{\rm ice}=8$--25 g\,cm$^{-2}$, in order to grow to $M_{\rm crit}$ before the gas disperses \citep{Inaba2003, Alibert2005, Hubickyj2005, Thommes2003}.

A core generates spiral density waves in the gas which cause the core to undergo inward type-I migration at a rate given by:
\begin{equation}
\left.\frac{da}{dt}\right)_{\rm I}\simeq-(2.7-1.1x)\left(\frac{M}{M_\star}\right)
\left(\frac{\Sigma_{\rm gas} a^2}{M_\star}\right)
\left(\frac{v_{\rm kep}}{c_s}\right)^2v_{\rm kep}
\label{eq-type1}
\end{equation}
where $v_{\rm kep}$ is the core's Keplerian orbital velocity, $c_s$ is the gas sound speed, and $M_\star$ is the stellar mass \citep{Tanaka2002}. (In this paper we consider only migration caused by the gaseous component of the disk). A $10M_\oplus$ core at 5 AU in a mass-enhanced nebula will migrate into its star in 10,000--30,000 years, much less than the time required to form a core in most models \citep{Inaba2003, Alibert2005, Hubickyj2005, Thommes2003, Ida2004}. Growth and migration rates are both proportional to $\Sigma$ (for a given disk metallicity), so this result is independent of the disk mass.

A massive planet clears an annular gap in the disk and undergoes type-II migration, moving inwards at the same rate that gas flows viscously towards the star:
\begin{equation}
\left.\frac{da}{dt}\right)_{\rm II}=-1.5\alpha\left(\frac{c_s}{v_{\rm kep}}
\right)^2v_{\rm kep}
\label{eq-type2}
\end{equation}
where $\alpha=\nu v_{\rm kep}/(ac_s^2)$ and $\nu$ is the disk viscosity  \citep{Dangelo2002}. 
Observed disk accretion rates imply $\alpha\sim 0.001$ \citep{Hueso2005}, so a planet at 5 AU will migrate into its star in 0.5 Myr. This is less than the lifetime of most disks \citep{Haisch2001}. 

\section{Living with Migration}

The existence of gas-giant planets suggests that: (i) giant-planet cores grow rapidly before type-I migration moves them into the star, and (ii) giant planets form at locations where the disk viscosity is lower than observed disk accretion rates would suggest.

Core growth should be fastest when planetesimals are small for two reasons. Small planetesimals experience strong gas drag, reducing $e$ and $i$ sufficiently for $v_{\rm rel}$ to be determined by Keplerian shear in the disk \citep{Rafikov2004, Chambers2006}. As a result, $P_{\rm col}$ is much higher than in the dispersion-dominated regime considered by most models. In addition, small planetesimals are slowed when they pass through the atmospheres of large cores, increasing the capture probability \citep{InabaIkoma}.

Most previous models have considered planetesimals with radii $r=10$--100 km \citep{Inaba2003, Alibert2005, Hubickyj2005, Thommes2003}. Planetesimals would have this size if they formed via gravitational instabilities (GI) in the solid component of the disk \citep{Wetherill1980}. However, GI requires $\Sigma_{\rm solid}/\Sigma_{\rm gas}\simeq 1$ \citep{Garaud2004}, which probably occurs under only limited circumstances. Alternatively, planetesimals may form by pairwise coagulation of dust grains \citep{Weidenschilling1997}, but it is unclear how large these objects will be. Regardless of the initial planetesimal size, $r$ probably becomes small once cores grow larger than Ceres, due to collisional fragmentation \citep{Kenyon2004}. Given these uncertainties, we will treat $r$ as a model parameter.

We can gauge the importance of $r$ using a simple model for core growth based on equation~\ref{eq-dmdt}. Here, $e$ and $i$ are determined assuming an equilibrium between excitation due to perturbations from nearby cores \citep{Ohtsuki2002} and damping due to gas drag \citep{Inaba2001}. The core's capture cross section is calculated including the effect of its atmosphere using equations A7--A12 of \cite{InabaIkoma}. Core growth rates are assumed to be 50\% higher than in equation~\ref{eq-dmdt} due to collisions between neighbouring cores \citep{Chambers2006}.

Figure~1 shows the mass of a core growing at 5 AU, neglecting migration, for 4 values of $r$. Here, $\Sigma_{\rm solid}=10$ g\,cm$^{-2}$, solids have a density of 1.5 gcm$^{-3}$, and $\Sigma_{\rm gas}/\Sigma_{\rm solid}=90$. Core growth times are shorter when $r$ is small, as expected. When $r\le 100$ m, the core formation time is shorter than the type-I migration timescale, shown by the dashed line. Dissipation of the nebular gas will slow type-I migration, and migration will speed up core growth by increasing the supply of planetesimals \citep{Alibert2005}. Hence, cores should survive when $r$ is somewhat larger than 100 m.

The source of viscosity in protoplanetary disks is unclear. One possibility is magneto-rotational instability \citep{Hawley1995}, which is highly effective but probably confined mainly to the surface layers of a disk \citep{Matsumura2006}. Near the midplane, where planets form, $\nu\sim 30\times$ lower than in the surface layers \citep{Turner2006}, corresponding to $\alpha\sim 3\times 10^{-5}$. Similar values of $\alpha$ can explain the size distribution  of chondrules seen in meteorites \citep{Cuzzi2001}. Low $\alpha$ also promotes dust coagulation by reducing typical collision speeds and fragmentation. For $\alpha=3\times 10^{-5}$, the type-II migration time for a planet at 5 AU  is $\sim 15$ Myr,  longer than the lifetime of most  disks \citep{Haisch2001}.

A core opens a gap in the disk when $M>M_{\rm gap}={\rm max}(M_{\rm nv},M_{\rm vis})$, where $M_{\rm nv}$ is the mass needed to open a gap in a non-viscous disk, given by
\begin{equation}
M_{\rm nv}\simeq\frac{2M_\star}{3}\left(\frac{c_s}{v_{\rm kep}}\right)^3
{\rm min}\left[5.2Q^{-5/7},\,3.8\left(\frac{c_s}{Qv_{\rm kep}}\right)^{5/13}\right]
\label{eq-nv}
\end{equation}
where $Q=c_sM_\star/v_{\rm kep}\Sigma_{\rm gas}\pi a^2$ is the Toomre stability criterion \citep{Rafikov2002}; and $M_{\rm vis}$ is the mass needed to maintain a gap against the viscous flow of the gas:
\begin{equation}
M_{\rm vis}\simeq M_\star\sqrt\frac{40\alpha\,c_s^5}{v_{\rm kep}^5}
\label{eq-vis}
\end{equation}
\citep{Lin1986}.
In the MMSN with $\alpha=3\times 10^{-5}$, $M_{\rm gap}\sim 15 M_\oplus$ at 5 AU. Numerical simulations show that gas flows onto a core even when a gap exists \citep{Dangelo2002}, allowing a gas-giant planet to form. When $M>M_{\rm gap}$, the maximum envelope growth rate is
\begin{equation}
\left.\frac{dM}{dt}\right)_{\rm gas,max}\simeq 2\pi a\Sigma_{\rm gas}\left|\left(\frac{da}{dt}\right)_{\rm II}
\right|\left[0.04+1.668\left(\frac{M}{M_J}\right)^{1/3}\exp{\left(-\frac{2M}{3M_J}\right)}\right]
\label{eq-gasmax}
\end{equation}
where $M_J$ is the mass of Jupiter \citep{Alibert2005, Veras2004}. In the MMSN with $\alpha=3\times 10^{-5}$, a $10M_\oplus$ core at 5 AU will grow to $M_J$ in $\sim 4$ Myr according to equation~\ref{eq-gasmax}.

\section{Simulations of Growth with Migration}

We now examine the growth and migration of giant planets using a more detailed model. We consider a disk with $2.5\le a\le 50$ AU, containing 640 cores, each of $10^{-4}M_\oplus$, and a population of planetesimals. Cores sweep up planetesimals following Eq.~\ref{eq-dmdt}. At each radial location, $e$ and $i$ vary due to gas drag and perturbations from the cores, calculated independently following \citet{Inaba2001} and \citet{Ohtsuki2002}. Cores are assumed to have circular, coplanar orbits due to dynamical friction with planetesimals and tidal interaction with the gas. Neighbouring cores merge when their orbital separation $<8r_h$, maintaining the typical spacing of 10--12 $r_h$ seen in N-body simulations of oligarchic growth \citep{Kokubo1998}. The capture cross section of a core's atmosphere is calculated following \cite{InabaIkoma}. Cores undergo migration according to Eqs.~\ref{eq-type1} and \ref{eq-type2}. The maximum type-II migration rate is determined by the rate of angular momentum transport through the disk \citep{Ida2004}. The core gap-opening mass is determined by Eqs.~\ref{eq-nv} and \ref{eq-vis}. 

The critical mass for a core to begin accreting a gaseous envelope is
\begin{equation}
M_{\rm crit}\simeq M_{c}\left(\frac{\dot{M}}{10^{-6} M_\oplus\ {\rm yr}^{-1}}\right)^{1/4}
\left(\frac{\kappa}{1\ {\rm cm}^2\ {\rm g}^{-1}}\right)^{1/4}
\label{eq-mcrit}
\end{equation}
where $\kappa$ is the opacity of the core's atmopshere, and $\dot{M}$ is the rate at which the core sweeps up planetesimals \citep{Inaba2003, Ikoma2000,Ida2004}. Here, we assume $\kappa=0.05$ cm$^2$\,g$^{-1}$ and $M_c=20M_\oplus$. When $M>M_{\rm crit}$, the envelope growth rate  is
\begin{equation}
\left.\frac{dM}{dt}\right)_{\rm gas}\simeq\left(\frac{M}{M_{\rm Earth}}\right)^3
\left(\frac{\kappa}{1\ {\rm cm}^2\,{\rm g}^{-1}}\right)^y
\frac{M}{10^9\ {\rm year}}
\end{equation}
\citep{Ikoma2000,Ida2004} where $y=-1/4$ is an empirical fit based on simulations by \cite{Hubickyj2005}. For cores that have opened a gap, the maximum gas accretion rate is given by Eq.~\ref{eq-gasmax}.  Initially, $\Sigma_{\rm solid}$ and $\Sigma_{\rm gas}$ are given by Eqs.~\ref{eq-sigma} with $x=-1/2$, modified by a factor of $\exp{(-a/25\ {\rm AU})}$. This profile is shallower than the MMSN to allow for migration and planetesimal drift due to gas drag. The disk is gravitationally stable everywhere such that $Q>2$. Initially $\Sigma_{\rm gas}/\Sigma_{\rm solid}=90$, and gas disperses exponentially over time with a time constant of 1 Myr. Solid material has a density of 1.5 g\,cm$^{-3}$, the stellar mass is $1 M_\odot$, and $r$ and $\alpha$ are assumed to be independent of $a$ and time.  Simulations last for 10 million years.

Figure~2 shows the evolution of three cores in a simulation with $r=100$ m, $\alpha=3\times 10^{-5}$, and $\Sigma_{\rm solid}=6$ gcm$^{-2}$ at 5 AU. The upper panel shows the cores' orbital evolution, with arrows showing when each core opens a gap. The core at 5 AU grows rapidly and begins noticeable type-I migration after $\sim 30,000$ years. At 50,000 years, the core opens a gap and type-II migration begins. However, inward migration is offset by the angular momentum gained by absorbing smaller bodies that migrate into the core's vicinity. These collisions can be seen as jumps in the core's mass, shown in the lower panel of Figure 2. After 40,000 years, the core starts to acquire an envelope. Envelope growth slows over time as $\Sigma_{\rm gas}$ decreases and $M$ increases, reducing accretion across the gap.  A second core at 20 AU undergoes rapid type-I migration, then opens a gap when $a\sim 9$ AU. This core acquires an envelope but its growth lags behind the body at 5 AU. The core moves inwards by $\sim 1$ AU due to type-II migration, but migration slows when $\Sigma_{\rm gas}$ becomes small. A third core at 23 AU grows and undergoes type-I migration, but remains too small to acquire an envelope before the gas disperses. These three planets are close analogues of Jupiter, Saturn and Uranus respectively.

Figure 3 shows the outcome of three simulations with different disk masses. Each row of symbols shows the surviving planets, with symbol radius $\propto M^{1/3}$. The black and grey symbol segments show the solid and gas mass fractions respectively. The numbers to the left of the symbols indicate $\Sigma_{\rm solid}$ at 5 AU. The last row of symbols shows the giant planets of the Solar System, where solid mass fractions include elements heavier than helium in each planet's envelope. More massive disks lead to larger final planets and allow giant planets to form further from the star. The model predicts that gas-giant planets will form and survive migration provided that $\alpha\le 3\times 10^{-4}$ and $r\le 0.5$ km. Most simulations generate radially ordered systems: 1 or 2 gas-giant planets with $3<a<20$ AU, followed by 1 or 2 large cores containing little gas, and finally a disk of sub-Earth-mass objects, akin to the Kuiper belt. 

Gas-giant planets produced in the model have solid-to-gas ratios similar to Saturn but higher than Jupiter. High solid fractions arise when the innermost planet absorbs other cores as they migrate inwards. In the model, closely spaced cores always coalesce. Jupiter's low solid-to-gas ratio suggests it gravitationally scattered nearby cores rather than absorbing them. In most simulations, several cores cross the disk's inner edge due to type-I migration. Presumably such objects would be lost into the star. These bodies contain only a few Earth masses and migrate rapidly, so their dynamical effect on nascent terrestrial planets is probably not severe. No planets are lost via type-II migration unless $\alpha\ge 10^{-4}$. A gas giant migrating through the inner Solar System would remove most of the solid material present. Terrestrial planets that formed subsequently would contain mostly ice-rich material from the outer disk \citep{Raymond2006}. The rocky compositions of the inner planets in the Solar System suggest that no giant planets were lost this way and that $\alpha$ was small.

If $\alpha\sim 10^{-4}$ at the disk midplane, type-II migration times will be comparable to disk lifetimes. In a flared disk, the migration rate is roughly independent of $a$ (see Eq.~\ref{eq-type2}). Hence, giant planets should be common everywhere between their formation location and the inner edge of the disk. The model predicts that this is the case. Figure 4 shows the planets produced in 48 simulations with $100\le r\le 500$ m and $3\times 10^{-5}\le \alpha\le 3\times 10^{-4}$. Giant planets are abundant at all distances from the inner edge of the disk out to 20 AU. This is in accord with the observed distribution of extrasolar planetary orbits \citep{Marcy2005}.
 
\acknowledgments

I would like to thank Alan Boss, Lindsey Chambers, Fred Ciesla, Roman Rafikov and an anonymous referee. This work was supported by NASA's TPF foundation science program.

\clearpage

\begin{figure}
\includegraphics[scale=.70,angle=90]{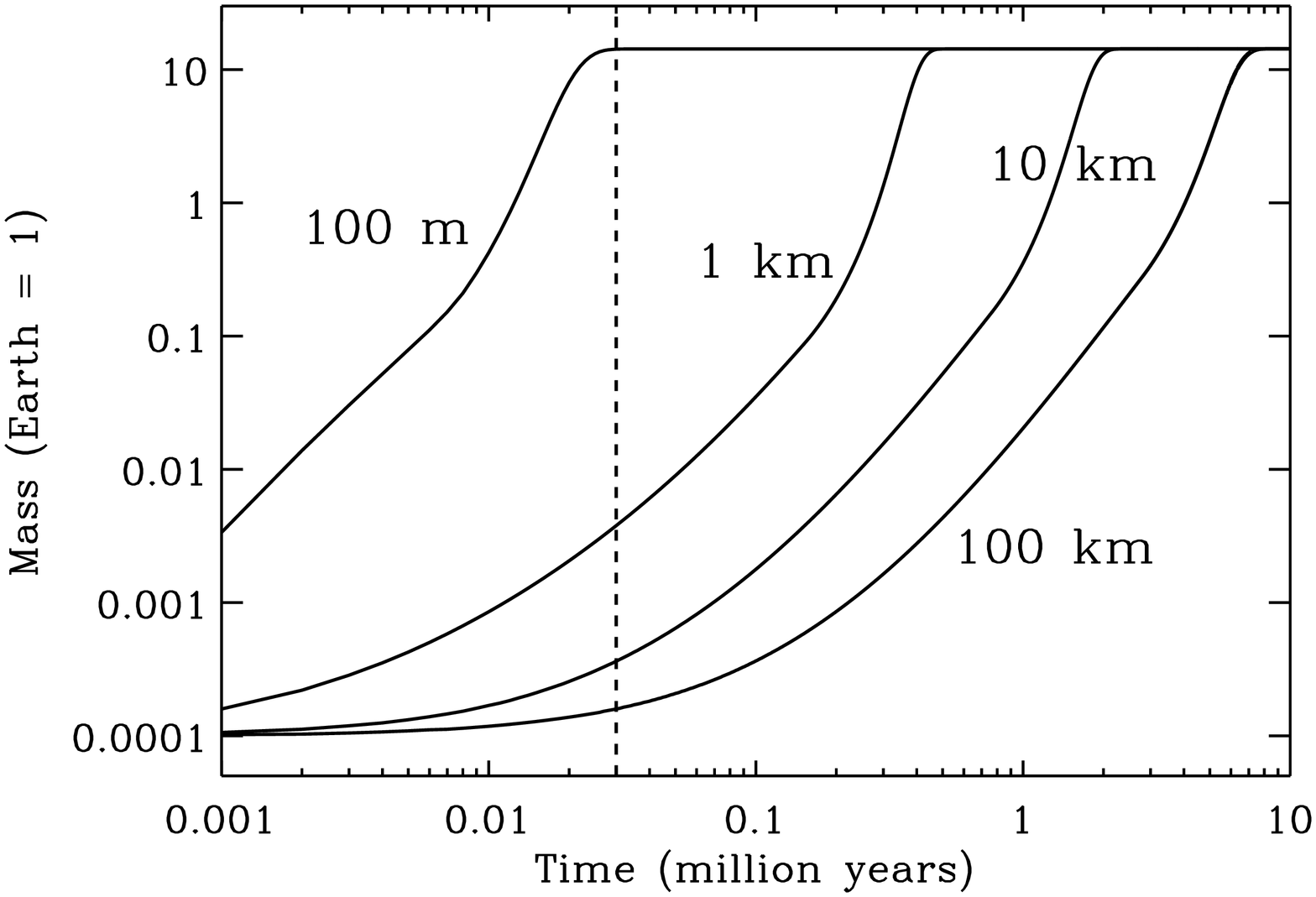}
\caption{Growth of a core at 5 AU from the Sun according to Eq.~\ref{eq-dmdt}, with $e$ and $i$ determined by an equilibrium between perturbations from nearby cores and damping due to gas drag. Migration is neglected. Here, $\Sigma_{\rm solid}$=10 g\,cm$^{-2}$ and $\Sigma_{\rm gas}/\Sigma_{\rm solid}=90$. Each curve shows growth for a different  $r$. The dotted line shows the type-I migration timescale.}
\end{figure}

\clearpage

\begin{figure}
\includegraphics[scale=.70]{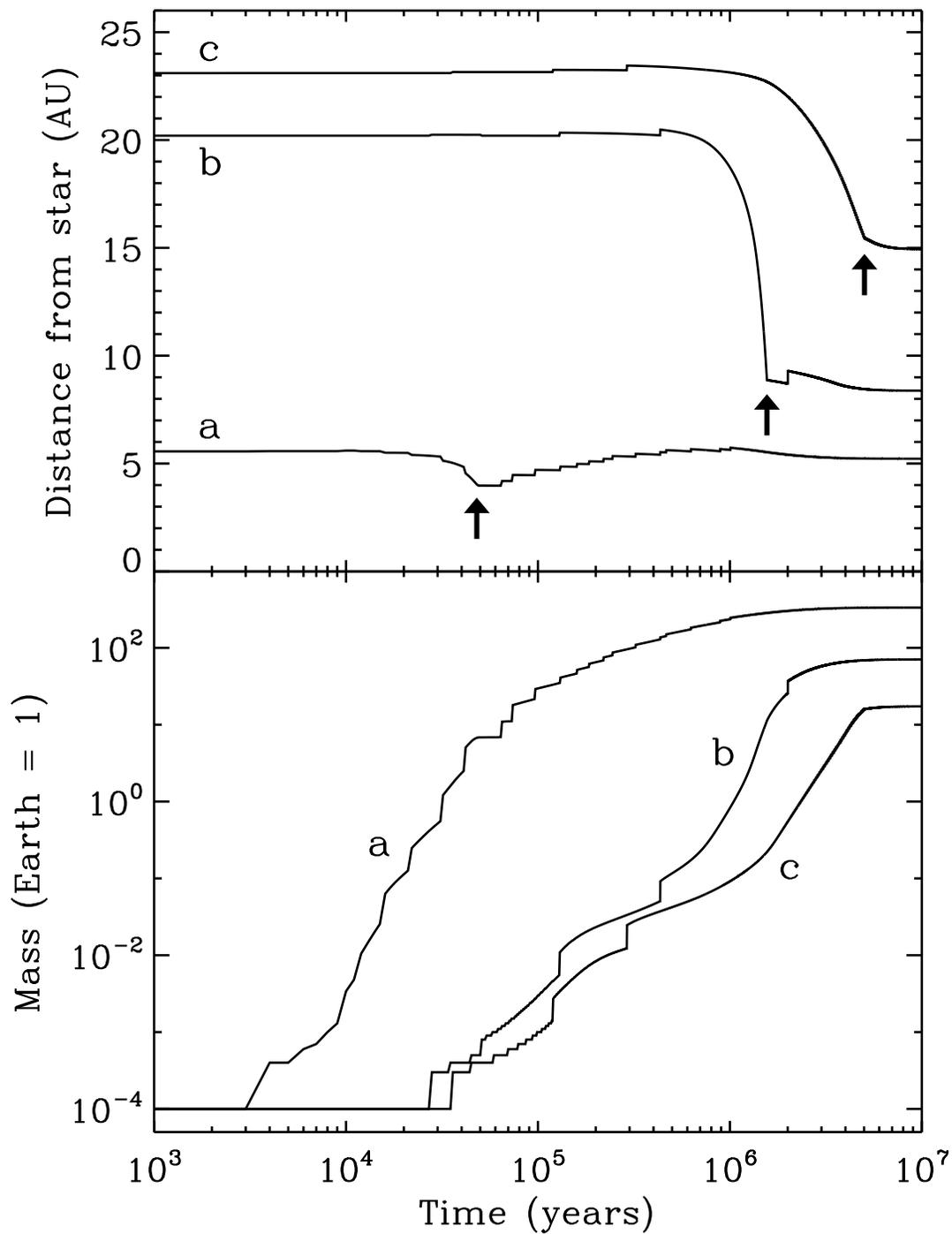}
\caption{Evolution of three surviving cores in a protoplanetary disk with $r=100$ m, $\alpha=3\times 10^{-5}$, and $\Sigma_{\rm solid}=6$ g\,cm$^{-2}$ at 5 AU. Growth is calculated using the model described in Section 3. Upper panel: distance from the star. Arrows show when each core opens a gap in the disk. Lower panel:  core masses.}
\end{figure}

\clearpage

\begin{figure}
\includegraphics[scale=.70,angle=90]{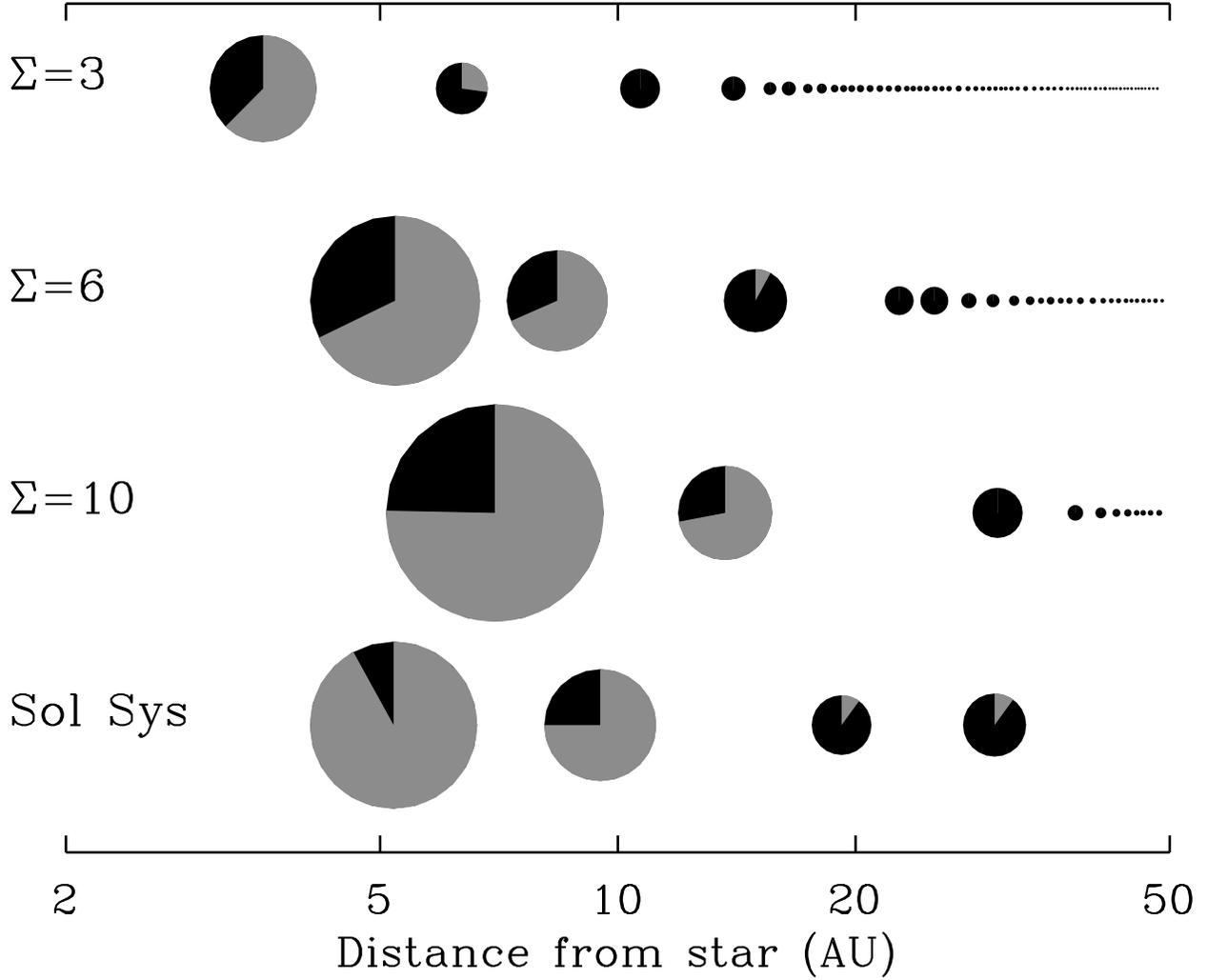}
\caption{The outcome of 3 simulations with $r=100$ m and $\alpha=3\times 10^{-5}$. The value of $\Sigma_{\rm solid}$ at 5 AU in each case is indicated. Surviving planets are represented by symbols with radius $\propto M^{1/3}$. The black and grey segments of each symbol give the solid and gas mass fractions respectively. The symbols marked `Sol Sys'  shows the giant planets of the Solar System.}
\end{figure}

\clearpage

\begin{figure}
\includegraphics[scale=.70,angle=90]{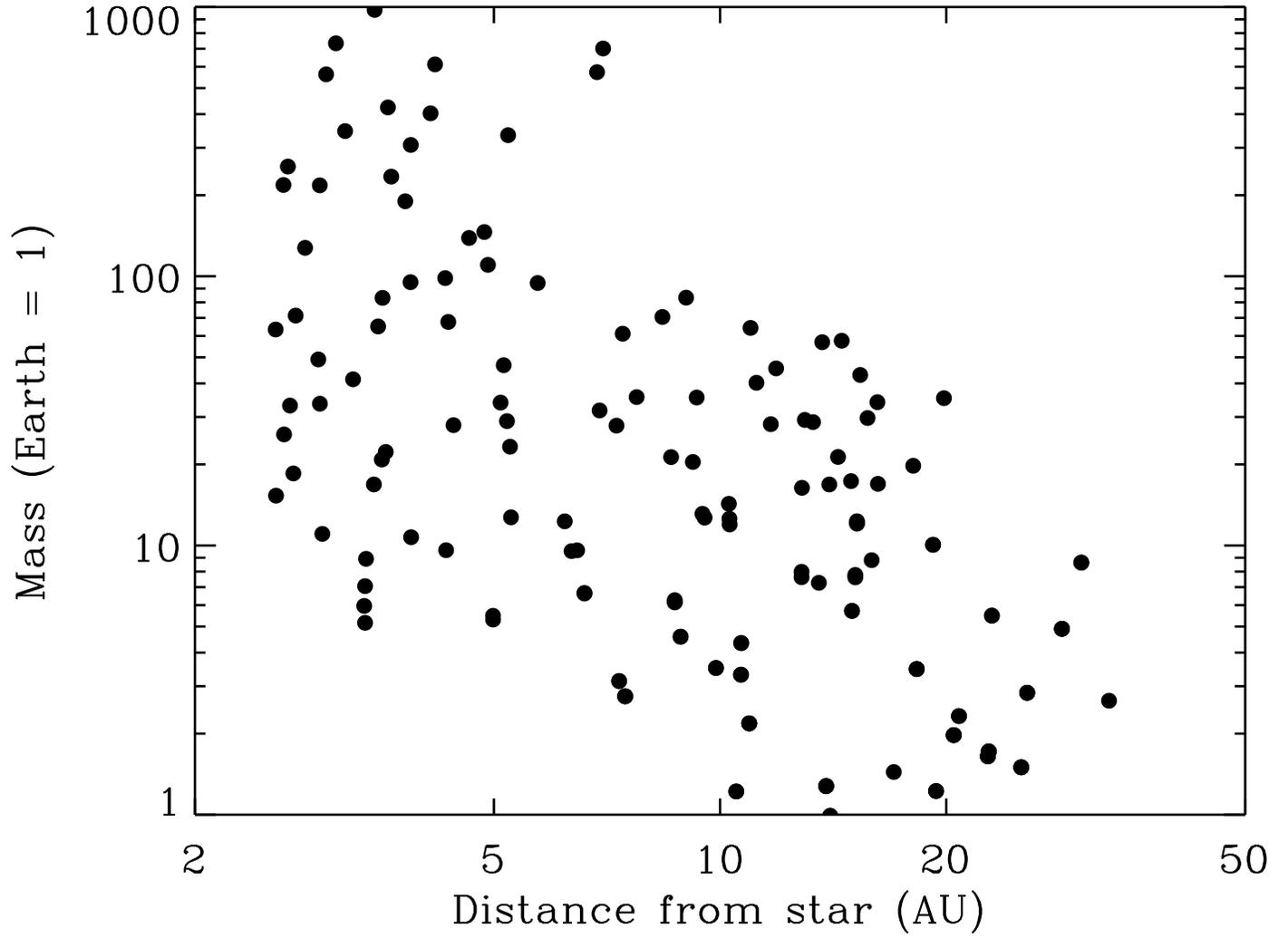}
\caption{Planets generated in 48 simulations with $100\le r\le 500$ m and $3\times 10^{-5}\le\alpha\le 3\times 10^{-4}$.}
\end{figure}

\end{document}